\documentclass[12pt]{iopart}
\usepackage{iopams}
\usepackage{latexsym} 
\usepackage{xcolor}
\usepackage{calrsfs}
\DeclareMathAlphabet{\pazocal}{OMS}{zplm}{m}{n}

\newcommand{\Sa}{\pazocal{S}}

\newtheorem{prop}{Proposition} 
\newtheorem{thm}{Theorem}
\newtheorem{lemma}{Lemma}
\newtheorem{cor}{Corollary}
\newtheorem{defn}{Definition}

\newtheorem{rmk}{Remark}

\newcommand{\Ad}{\mbox{${\rm Ad}$}}
\newcommand{\epf}{\hfill \mbox{$\Box $}}

\newcommand{\rf}{\mbox{${\mathbb R}$}}

\begin{document}

\title[Discrete nonlinear Fourier transforms and their inverses]
{Discrete nonlinear Fourier transforms and their inverses}

\author{Pavle Saksida}
\address{Faculty of Mathematics and Physics, 
University of Ljubljana, Jadranska 21, 
1000 Ljubljana, Slovenia}

\ead{Pavle.Saksida@fmf.uni-lj.si}

\begin{abstract}
We study two discretisations of the nonlinear Fourier transform of
AKNS-ZS type, ${\cal F}^E$ and ${\cal F}^D$. 
Transformation ${\cal F}^D$
is suitable for studying the distributions of the form $u = \sum_{n = 1}^N u_n \, \delta_{x_n}$,
where $\delta _{x_n}$ are delta functions. The poles $x_n$  are not equidistant.
The central result of the paper is the construction of recursive algorithms for inverses
of these two transformations.  The algorithm
for $({\cal F}^D)^{- 1}$ is numerically more demanding than that for
$({\cal F}^E)^{- 1}$.
We describe
an important  symmetry property of ${\cal F}^D$. It enables the reduction of the nonlinear Fourier
analysis of  the constant mass distributions $u = \sum_{n = 1}^N  u_c \, \delta _{x_n}$ for the numerically 
more efficient ${\cal F}^E$ and its inverse.
\end{abstract}


\maketitle

\section{Introduction}

The theory of nonlinear Fourier analysis stems from the inverse scattering
method for solving nonlinear integrable partial differential equations. The inverse
scattering method is a nonlinear version of the original Fourier's idea for solving
the initial value problems for linear partial differential equations.
The Fourier approach first uses the Fourier transform to recast the initial condition
$u(x, 0)$ into data whose time evolution is easily found. We find the transformed data for future time $t$
and then apply the inverse Fourier transform to find the value of the solution $u(x, t)$ of our
problem at time $t$. Roughly speaking, the inverse scattering method has the same basic structure, except that the linear Fourier transform has to be replaced by a nonlinear one and the
inverse linear transform by its nonlinear analogue. 
Therefore, finding the inverse nonlinear Fourier transform is an important problem.  The inverse scattering method
approach was discovered by Gardner, Greene, Kruskal and Miura in \cite{GGKM1},
\cite{GGKM2}, where they solved the Korteweg-deVries equation.
There are many versions of the nonlinear Fourier transforms. One of the most popular is the transform,
associated with the AKNS-ZS systems. This was introduced in the pioneering work of
Ablowitz, Kaup, Newell, and Segur, and simultaneously by Zakharov and Shabat, see \cite{AKNS1},
\cite{AKNS2}, and \cite{ZS}. 

In this study, we consider two discretisations of this
transform and their inverses for functions, defined on a finite interval. The first discretisation
${\cal F}^E$ is a  simple Euler type construction.  The second
discretisation ${\cal F}^D$ is more interesting. It  can be used to transform the finite linear combinations of delta functions
of the form $u = \sum_{n = 1}^N u_n \, \delta _{x_n}$. Here, the points $x_n$ are not presumed to be
equidistant. The arrays of
values $\{u_n\}_{n = 1, \ldots, N}$ and $\{ x_n \}_{n = 1, \ldots, N}$ are both
the unknowns of the inverse of ${\cal F}^D$.
Distributions of the form $\sum u_n \, \delta _{n_n}$  are of great interest in many fields of mathematics and their applications
especially for signal processing. Crystalline measures and one-dimensional Fourier quasicrystals are special cases of infinite versions of such distributions. For an exciting work on these objects, see \cite{KuSa}, \cite{Meyer}, \cite{LevOlevinskii},
\cite{OlevskiiUlanovskii}, to mention but a few. The main tool of the experimental work in this field is
finite sums of the form $u =\sum_{n = 1}^N u_n \, \delta _{x_n}$.
The transform ${\cal F}^D$
can also be used as an approximation of the nonlinear transform of continuous functions when
we apply it to the distributions of the form $u = \sum_{n = 1}^N u_n \, \Delta x_n \delta_{x_n}$,
where $\Delta x_n = x_{n -1} - x_n$.

The main
results of this study are Theorems \ref{thmD} and \ref{Ethm}. They provide 
recursive algorithms for the evaluation of the inverse transformations of ${\cal F}^D$ and
${\cal F}^E$. Not surprisingly, the algorithm for inverse $({\cal F}^E)^{- 1}$
is numerically much cheaper than the one for $({\cal F}^D)^{- 1}$. In
\cite{NLFT-1} and \cite{NLFT-2} we provided perturbative constructions for the
evaluations of two types of continuous AKNS-ZS nonlinear Fourier transforms.
Our algorithms in this study are recursive, but they are not perturbative. In principle, they yield the exact
solutions to the two inverse problems in finite time.

Transforms ${\cal F}^E$ and ${\cal F}^D$ send functions or distributions into functions
$z \mapsto {\cal A}(z)$, where ${\cal A}(z)$ are matrices of the form
\[
{\cal A}(z) = \pmatrix{ a(z) & b(z) \cr
                                   - \overline{b(z)} & \overline{a(z)} }.
                                   \] 
All  such functions are not elements of the images of ${\cal F}^E$ or ${\cal F}^D$.
As corollaries of theorems \ref{thmD} and \ref{Ethm}, we give the tests which decide when
a matrix function ${\cal A}(z)$ is in the image of ${\cal F}^E$ or  ${\cal F}^D$.

Suppose we know in advance that for ${\cal A}(z) ={\cal F}^D[v]$ for some distribution
$v = \sum_{n = 1}^N v_n \, \delta_{(n \Delta x)}$, whose poles are equidistant.
Then it is not difficult to see that one can essentially use the faster method
for $({\cal F}^E)^{- 1}$ to determine the unknown $v = ({\cal F}^D)^{- 1}[{\cal A}]$.
The question arises whether this can also be done in the case where the distributions in question are the constant mass distributions, that is, those of the form 
$w = \sum_{n = 1}^N  u_c \, \delta _{x_n}$.
The answer to this question is negative. However, in ${\cal F}^D$, the roles of the variables $u_n$ and $\Delta x_n = x_{n + 1} - x_n$ are almost symmetrical.
Reversing the roles of $u_n$ and $\Delta x_n$ yields the nonlinear Fourier transformation ${\cal F}^D_d$ which can be considered dual to
${\cal F}^D$, as explained in Section \ref{dualitysec}. 
Theorem \ref{dual} shows the following. Let ${\cal B}(z)$ be equal
to ${\cal F}_d^D[w]$ for a constant mass distribution $w$ with non-equidistant poles. Then
$({\cal F}^D_d)^{- 1}[{\cal B}]$ can be computed using a faster algorithm
for the evaluation of $({\cal F}^E)^{- 1}$.

We hope our results will contribute some new insight to the developing field of nonlinear Fourier analysis.
In particular, the transformation ${\cal F}^D$ could provide a helpful tool for the nonlinear Fourier
analysis of distributions.
Nonlinear Fourier analysis is becoming an essential part
of the field of harmonic analysis and its applications in a broad sense; see the references 
\cite{DauxPey}, \cite{Osborne},  \cite{LaserRadiation}, \cite{TTIntro}, \cite{TTM1}, \cite{TTM2}, 
to name but a few. The applicability of
the linear Fourier transform is enormous, and the relation between the linear and nonlinear Fourier transforms is better
understood in time, see the references \cite{FoGe}, \cite{Pe}, \cite{YK1}, \cite{YK2}, and many other
works. Undoubtedly, the role of nonlinear Fourier analysis will become ever more critical in studying various nonlinear problems.

The second section recalls the definition and basic properties of AKNS-ZS nonlinear Fourier
transform for functions in finite intervals. In Section 3, we introduce  discretisations ${\cal F}^E$
and ${\cal F}^D$, respectively. The central part of the paper is sections 4 and 5, in which we construct the
algorithms for evaluating $({\cal F}^D)^{- 1}$ and $({\cal F}^E)^{- 1}$, respectively. In section 6,
we describe the symmetric structure of ${\cal F}^D$ and introduce the dual transformation
${\cal F}^D_d$. We apply it to the study of the constant mass distributions of the form
$w = \sum_{n = 1}^N u_c \, \delta _{x_n}$. We conclude the paper with some brief remarks
on the computational complexity of our algorithms. We also mention possible directions for
further research.

\section{Nonlinear Fourier transform}
Let $LSU(2)$ denote the space of
functions of a real variable $z$, with values in the group of $SU(2)$ matrices given by
\[
z \longmapsto \pmatrix{ a(z) & b(z) \cr
- \overline{b(z)} & \overline{a(z)} },
\]
where the functions $a(z)$ and $b(z)$ are elements of the $L^2$-space on a finite interval,
say $[0, 1]$. Let $x \mapsto u(x)$ be an $L^2[0, 1]$ complex-valued function.
The nonlinear Fourier transform associated with the AKNS-ZS integrable systems on  a finite interval
 is the map
\[
{\cal F} : L^2[0, 1] \longrightarrow LSU(2), \quad u(x) \longmapsto {\cal F}[u](z)
\]
given by the following rule: Let $\Phi[u](x; z)$ be the solution of the family of linear initial value
problems
\begin{equation}
\Phi_x(x; z) = L(x; z) \cdot \Phi(x; z), \quad \Phi(0; z) = I, \quad {\rm for} \ {\rm every} \ z,
\label{initial}
\end{equation}
where the coeficient matrix is given by
\[
L(x, z) = \pmatrix{ \pi i z & u(x) \cr
- \overline{u(x)} & - \pi i z }.
\]
Then,
\[
{\cal F}[u](z) = \Phi(x = 1; z).
\]
The above initial value problem can be solved by a convergent function
series, namely the Dyson series. We have
\[
\Phi(x;z) = I + \sum_{d = 1}^{\infty} \int_{\Delta _d(x)} L(x_d, z) \cdot L(x_{d - 1}, z) \cdots
L(x_1, z) \ d\vec{x}.
\]
Above $\Delta_d(x)$ denotes the ordered $d$ dimensional simplex
\[
\Delta _d(x) = \{(x_1, x_2, \ldots, x_d) \in \rf^d; 0\leq x_1 \leq x_2 \leq \ldots \leq x_d \leq x \}.
\]
The linearisation of the nonlinear Fourier transform ${\cal F}[u]$ around $u \equiv 0$ is the
usual linear Fourier transform. To see this, it is helpful to change the gauge using the gauge
transformation matrix $G(x, z) = {\rm diag}( e^{- \pi i x z}, e^{\pi i x z})$. The transformed
matrix $L^G$ of coefficients is given by
\[
L^G = G_x \cdot G^{- 1} + G \cdot L \cdot G^{- 1}
\]
or explicitly
\[
L^G(x, z) = \pmatrix{ 0 & e^{- 2 \pi i x z} u(x) \cr
- e^{ 2 \pi i x z} \overline{u(x)} & 0 }.
\]
The nonlinear Fourier transformation ${\cal F}^G$ in the new gauge is given by
\[
{\cal F}^G[u](z) = I + \pmatrix{ 0 & F[u](z) \cr
- \overline{F[u](z)} & 0 } + \sum_{d = 2}^{\infty}A_d[u](z).
\]
Here, $F[u](z)$ is the usual linear Fourier transform of $u$, and $A_d[u]$ are operators of order $d$ in $u$.
We have
$
{\cal F}^G[u](z) = G(1, z) \cdot {\cal F}[u](z)
$. 
When $z$ is taken to be an integer-valued variable, we have
${\cal F}^G [u](n) = (- 1)^n {\cal F}[u](n)$.

\section{Discretizations of nonlinear Fourier transform}

We shall describe two simple discretizations of the nonlinear Fourier transform.

\paragraph{\it \underline{Euler type of discretization:}} In the initial value problem (\ref{initial}) we can
replace the differential equation using a difference equation. We replace the function $u(x)$ with a
function of the discrete variable $n \mapsto u_n, \ n = 0,\ldots, N - 1$, for some $N \in \mathbb{Z}$, and obtain
\[
\frac{\Phi(n + 1, z) - \Phi(n, z)}{\frac{1}{N}} = L^G(n, z) \cdot \Phi(n, z), \quad \Phi(0, z) = I
\]
for
\begin{equation} \fl 
L^G(n, z) = \pmatrix{0 & e^{ - \frac{2 \pi i n z}{N}} \ u_n \cr
- e^{ \frac{2 \pi i n z}{N}} \ \overline{u_n} & 0 } = 
\pmatrix{ e^{ - \frac{2 \pi i n z}{N}} & 0 \cr
0 & e^{ \frac{2 \pi i n z}{N}} } \cdot 
\pmatrix{ 0 & u \cr
- u & 0 }.
\label{Ecoeffmatrix}
\end{equation}
This gives
\[
\Phi(n + 1, z) = (I + \frac{1}{N} L^G(n, z)) \Phi(n, z)
\]
and by recursion
\begin{equation}
{\cal F}^E[u](z) = \prod_{n = N - 1}^0 \Bigl( I + \frac{1}{N} L^G(n, z)\Bigr).
\label{EulerTransfProd}
\end{equation}
We restrict the values of $z$ to $\{0, 1, \ldots, N - 1\}$.
Matrix multiplication yields the polynomial analogue of the Dyson series, namely,
\begin{equation} \fl
{\cal F}^E[u](z) = I + \sum_{d = 1}^N \Bigl( \frac{1}{N^d}
\sum_{N - 1 \geq n_d >\ldots > n_2 > n_1 \geq 0}
\! \! \! \! \! \! L^G(n_d, z) \cdots L^G(n_2, z) \cdot L^G(n_1, z) \Bigr).
\label{EulerF1}
\end{equation}
Let us denote
\begin{equation}
E_N(n, z) = \pmatrix{ e^{ \frac{\pi i n z}{N}} & 0 \cr
0 & e^{ - \frac{\pi i n z}{N}} }, \quad \quad
U_m = \pmatrix{ 0 & u_m \cr
- \overline{u_m} & 0 }.
\label{EandJ}
\end{equation}
We have
\begin{equation}
U_m \cdot E_N(n, z) = E_N( - n, z) \cdot U_m.
\label{commutrel}
\end{equation}
Using (\ref{Ecoeffmatrix}) and (\ref{commutrel}),
we collect all the exponential factors $E_N(- n, z)$ on the left. Then, (\ref{EulerF1})
becomes
\begin{eqnarray} \fl \nonumber
{\cal F}^E[u](z) = I & +& \sum_{n = 0}^{N - 1} \frac{1}{N} \, E_N( - 2 n, z) \, \cdot U_n \\
& + &
\sum_{d = 2}^N \ \sum_{ N \geq n_d > \ldots, > n_1 \geq 0} \! \! (\frac{1}{N})^ d \
E_N\Bigl(- 2 \sum_{k = d}^1 (- 1)^{d - k} n_k , z\Bigr) \, \prod _{k = d}^1 U_{n_k}.
\label{EulerF2}
\end{eqnarray}
Then the linear part of (\ref{EulerF2}) is essentially the usual linear
discrete Fourier transform.
A similar approach was used in the continuous case in \cite{Fr-et-al}.

\paragraph{\it \underline{ Transform of a sum of delta functions:}}
Here, the spectral variable $z$ takes value in the entire $\rf$.
One of the most important ways to represent signals is to assign a suitable distribution of
the form
\begin{equation}
u = \sum_{n = 1}^N u_n \, \delta_{x_n}.
\label{sparsedelta}
\end{equation}
Let us introduce two discrete functions.
\begin{eqnarray}
\tilde{x} : \ n \longmapsto x_n, \quad n = 1, \ldots , N , \quad 0 < x_1 < x_2 < \ldots < x_N < 1,\\
\tilde{u }: \ n \longmapsto u_n, \quad n = 1, \ldots , N.
\end{eqnarray}
The values of $u_n$ are arbitrary complex numbers. The points $x_n$ are {\it not} assumed to be equidistant.
We shall define $\{\Delta x_n\}_{n = 0, \ldots, N}$ by
\[\fl
\Delta x_0 = x_1, \quad \quad
\Delta x_n = x_{n + 1}- x_n \ \ {\rm for} \ \ n = 1, \ldots , N - 1, \quad \quad \Delta x_N = 1 - x_N.
\]

Let now $u_{\epsilon}(x)$ be the step function, given by
\begin{equation}
u_\epsilon(x) = \left\{ \begin{array}{rl} \frac{1}{\epsilon}\,
u_n, & x \in [x_n - \frac{\epsilon}{2}, x_n + \frac{\epsilon}{2}], \ \ n = 1, \ldots , N\\
0, & {\rm otherwise}
\end{array} \right. .
\label{spikestep}
\end{equation}
Here, $\epsilon$ is a number that is smaller than the smallest $\Delta x_n$. So, the function
$u_\epsilon(x)$ is equal to zero at intervals $[0, x_1 - \frac{\epsilon}{2})$,
$(x_n + \frac{\epsilon}{2}, x_{n + 1} - \frac{\epsilon}{2})$ for $n = 1, \ldots, N$, and 
$(x_n + \frac{\epsilon}{2}, 1]$. At the intervals $[x_n - \frac{\epsilon}{2}, x_n + \frac{\epsilon}{2}]$, it assumes large values of  $\frac{ u_n}{\epsilon}$. In terms
of the theory of distributions, we have
\[
\lim _{\epsilon \to 0} u_{\epsilon}(x) = \sum _{n = 1}^{N} u_n \, \delta_{x_n}(x).
\]
The nonlinear Fourier transform, defined by (\ref{initial}),  of a step function $u_{\epsilon}$, is the product of the matrix exponentiations
of constant pieces of $u_{\epsilon}$. We have
\begin{equation}
 \fl
{\cal F}[u_\epsilon](z) = {\cal E}_N(\epsilon, z) \cdot {\cal R}_N(\epsilon, z) \cdots 
{\cal E}_2(\epsilon, z) \cdot
{\cal R}_2(\epsilon, z) \cdot {\cal E}_1(\epsilon, z) \cdot {\cal R}_1(\epsilon, z) \cdot 
{\cal E}_0(\epsilon, z),
\label{D-approx}
\end{equation}
where
\[ \fl
{\cal E}_n(\epsilon, z) = {\rm Exp}\Bigl((\Delta x_n - \widetilde{\epsilon}) \pmatrix{ \pi i z & 0 \cr
0 & - \pi i z } \Bigr)
, \quad n = 0, \ldots , N
\]
and
\[ \fl
{\cal R}_n (\epsilon, z) =
{\rm Exp}( \epsilon \pmatrix{ \pi i z & \frac{1}{\epsilon}\, u_n \cr
- \frac{1}{\epsilon} \, \overline{ u_n} & - \pi i z } ) =
{\rm Exp}(\pmatrix{ \epsilon \, \pi i z & u_n \cr
- \overline{u_n} & - \epsilon \, \pi i z }).
\]
In the definition of ${\cal E}_n$, we have $\widetilde{\epsilon} = \epsilon $ for $n = 1, \ldots , N - 1$
and $\widetilde{\epsilon} = \frac{\epsilon}{2}$ for $n = 0, \, N$.
A straightforward calculation gives
\[ \fl
{\cal R}_n(\epsilon, z) =
\pmatrix{ \cos{(A(u_n, \epsilon z))} + i \pi \epsilon z \, {\rm sinc}(A(u_n, \epsilon z)) & 
u_n \, {\rm sinc}(A(u_n, \epsilon z)) \cr
\! \! \! \! - u_n \, {\rm sinc}(A(u_n, \epsilon z)) & \! \! \! \! \! \! \! \! \!
\cos{(A(u_n, \epsilon z))} - i \pi \epsilon z \, {\rm sinc}(A(u_n, \epsilon z)) },
\]
where
\[
A(u_n, \epsilon \, z) = \sqrt{| u_n|^2 + (\epsilon \pi z)^2}.
\]
In the limit $\epsilon \to 0$, we have
\begin{equation}
\lim _{\epsilon \to 0} \, {\cal E}_n(\epsilon, z) = E(\Delta x_n, z) =
\pmatrix{ e^{\pi i \Delta x_n z } & 0 \cr
0 & e^{ - \pi i \Delta x_n z} }.
\label{E-limit}
\end{equation}
If we rewrite $u_n$ in polar coordinates,
$
u_n = e^{i \phi _n} r_n
$,
we get
\begin{equation}
\lim_{\epsilon \to 0}\, {\cal R}_n(\epsilon, z) = R( u_n) =
\pmatrix{\cos{( r_n)} & e^{i\phi _n} \sin{( r_n)} \cr
- e^{- i \phi_n} \sin{( r_n)} & \cos{ ( r_n )} }.
\label{R-limit}
\end{equation}

\begin{defn} Let the distribution $u$ be given by
$u = \sum_{n = 1}^N u_n \, \delta_{x_n}$, where $ 0 < x_1 < x_2 < \ldots < x_N < 1$.
Then the nonlinear Fourier transform ${\cal F}^D[u]$ of $u$ is given by
\[
{\cal F}^D[u](z) = \lim _{\epsilon \to 0} {\cal F}[u_{\epsilon}](z).
\]
\label{FD}
\end{defn}
The above calculations provide proof of the following proposition.

\begin{prop} Let 
$
u = \sum_{n = 1}^N u_n \, \delta_{x_n},
$
where $0 < x_1 < x_2 < \ldots < x_N < 1$, and $u_1, \ldots, n_n$ are arbitrary complex numbers.
The nonlinear Fourier transform of $u(x)$ is given by
\begin{equation}
{\cal F}^D[u](z) = \Bigl( \prod _{n = N}^1E(\Delta x_n, z) \cdot R( u_n).
\Bigr) \cdot E(\Delta x_0, z).
\label{firstdelta}
\end{equation}
\end{prop}
{\bf Proof:}
The proposition follows immediately from formulae (\ref{D-approx}) and (\ref{E-limit}), 
(\ref{R-limit}).

\epf
\noindent

\noindent Rewriting the formula for ${\cal F}^D[u]$ in terms of the variables $x_n$ instead of
$\Delta x_n$ yields the following expression:

\begin{cor}
The nonlinear Fourier transform of $u= \sum_{n = 1}^N u_n \, \delta _{x_n}$ can be expressed  as
the formula
\begin{eqnarray} \nonumber
{\cal F}^D[u](z) & = & E(1, z) \cdot \prod_{n = N}^1 E(- x_n, z)
\cdot R( u_n) \cdot E(x_n, z) \\
& = & E(1, z) \cdot \prod_{n = N}^1 \ {\rm Ad}_{E(- x_n, z) } R( u_n).
\label{AdRep1}
\end{eqnarray}
More explicitly,
\begin{equation} \fl
{\cal F}^D[u](z) =
\pmatrix{e^{\pi i z} & 0 \cr
0 & e^{-\pi i z} } \cdot
\prod_{n = N}^1 \pmatrix{ \cos{( r_n)} & e^{- 2 \pi i x_n z} \ e^{i \phi _n}\sin{( r_n)} \cr
- e^{2 \pi i x_n z} \ e^{ - i \phi_n}\sin{(r_n)} & \cos{( r_n)} }.
\label{seconddelta}
\end{equation}
\label{cor1}
\end{cor}
\epf

\noindent For easier writing, it will be handy to introduce a slightly modified, ``reduced'' transform
\begin{equation}
{\cal F}_r^D[u](z) = E(- 1, z) \cdot {\cal F}^D[u](z) = \prod_{n = N}^1 \Ad_{E(- x_n, z)} R(u_n).
\label{FDAd}
\end{equation}

At first glance, formula (\ref{firstdelta}) appears to be different from the Euler-type discretisation
${\cal F}^E$
of ${\cal F}$. However, (\ref{seconddelta}) enables us to find the analogue of Dyson's polynomial for
${\cal F}^D[u]$,
which makes the relation to the Euler discretization clearer.
\label{DtoE}
Let us introduce the matrix
\[
{\cal U}_n = \pmatrix{ 0 & e^{ i \phi_{n}} \sin{r_{n}} \cr
- e^{- i\phi_{n}} \sin{r_{n}} & 0 }.
\]
We have
\[ \fl
\pmatrix{ \cos{ r_n} & e^{- 2 \pi i x_n z} \ e^{i \phi _n}\sin{ r_n} \cr
- e^{2 \pi i x_n z} \ e^{ - i \phi_n}\sin{r_n} & \cos{ r_n} }
= \cos{r_n} \cdot I + E(- 2 x_n, z) \cdot {\cal U}_ n.
\]
Denote
\[ \fl
{\cal V}_{n_1, n_2, \ldots , n_d}(u) = (\prod _{m \ne n_k; k = 1, \ldots , d} \! \! \! \! \! \cos{r_m} )
\prod_{k = d}^1 {\cal U}_{n_k}, \quad \quad {\rm and} \quad \quad 
{\cal C}(u) = \prod_{n = 1}^N \cos{r_n}.
\]
Using the commutation rule
\[
E(x_n, z) \cdot {\cal U}_n = {\cal U}_n \cdot E(- x_n, z),
\]
the reduced (\ref{FDAd}) can be rewritten in the form
\begin{eqnarray} \fl \nonumber
{\cal F}_r^D[u](z) = {\cal C}(u) \, I & + & \sum_{n = 1}^{N} E(- 2 x_n, z) \cdot 
{\cal V}_n \\
& + & \sum_{d = 2}^N \sum_{N \geq n_d > \ldots n_2 > n_1 \geq 1} \! \! \! \!
E( - 2 \sum_{k = 1}^d (- 1)^{d - k} x_k, z ) \cdot {\cal V}_{n_1, n_2, \ldots, n_d}(u).
\label{thirddelta}
\end{eqnarray}
The expressions (\ref{EulerF2}) and (\ref{thirddelta}) are indeed similar. 
In particular, we note that expanding ${\cal V}_n(u)$ into the Taylor series around zero,
with respect to $u$ and keeping only the linear terms in the second summand 
\[ \fl
\sum_{n=1}^N E(- 2 x_n, z) \cdot {\cal V}_n (u) = 
\sum_{n = 1}^N 
\pmatrix{ 0 & \! \! \! \! \! \! e^{- 2 \pi i x_n z} \sin{u_n} \prod_{m \ne n} \cos{u_m} \cr
e^{- 2 \pi i x_n z} \sin{u_n} \prod_{m \ne n} \cos{u_m} & 0 }
\]
of (\ref{thirddelta})
yields the 
usual linear Fourier transform $ \sum_{n = 1}^N    u_n \, e^{- 2 \pi i x_n z} $ of 
$u = \sum_{n = 1}^N u_n \, \delta_{x_n}$.

Let us denote
$\widehat{u} = \sum_{n = 1}^N \Delta x_n u_n \, \delta_{x_n}$. The expression
${\cal F}^D[\widehat{u}](z)$
can be thought of as a nonlinear Fourier transform of the step function
\[
v(x) = \left\{ \begin{array}{rl} u_n, & x \in [x_{n - 1} , x_n ], \ \ n = 1, \ldots , N \\
0, & {\rm otherwise}
\end{array} . \right.
\]
Indeed, we consider ${\cal F}^D[\widehat{u}] / {{\cal C}(u)}$. Expand  each factor of the product
$\frac{{\cal V}_{n_1, n_2, \ldots , n_d}}{{\cal C}(u)}  = 
\prod_{k = 1}^d \tan{( \Delta{x_{n_k}}u_{n_k})}$ into Taylor series around zero with respect to the
variables $\Delta x_n u_n$ and keep only the linear terms. 
Let us, further, set $x_n = \frac{n}{N}$ for all $n$.
Then, also $\Delta x_n = \frac{1}{N}$, and the expression
(\ref{thirddelta}) becomes (\ref{EulerF2}).

\section{The inverse of ${\cal F}^D$}
\label{InvD}

In this section, we describe an algorithm for  evaluating the inverse of the nonlinear Fourier
transform ${\cal F}^D$. The following theorem gives our result:

\begin{thm}
Suppose we are given an $SU(2)$-valued function
\[
z \longmapsto {\cal A}(z) = \pmatrix{ \alpha (z) & \beta (z) \cr
- \overline{\beta(z)} & \overline{\alpha(z)} }
\]
such that $ {\cal A}(z)= {\cal F}_r^D[u](z)$, for some
$
u = \sum_{n = 1}^N u_n \,\delta _{x_n}
$ with $|u_n| < \frac{\pi}{2}$.
The array
\[{\cal U} =\{(x_1, u_1), (x_2, u_2), \ldots , (x_N, u_N)\}
\]
can be computed from ${\cal A}(z)$ by the following recursion.

\noindent Set ${\cal A}_0(z) = {\cal A}(z)$.

\noindent The matrix function ${\cal A}_{k + 1}(z)$ is obtained from ${\cal A}_k(z)$,
where
\[
{\cal A}_k(z) = \pmatrix{ \alpha_k(z) & \beta_k(z) \cr
- \overline{\beta_k(z)} & \overline{\alpha_k(z)} }
\]
by the following construction.
\begin{enumerate}
\item Compute the linear inverse Fourier transforms $F^{- 1}[\alpha_k]$ and $F^{- 1}[\beta_k]$.
\item From the results read-off the right-most delta term $b_k\, \delta_{y_k}$ of $F^{- 1}[\beta_k]$ and
the left-most term $a_k \, \delta_0$ of $F^{- 1}[\alpha]$.
\item Set
\[
x_{N - k} : = y_k, \quad {\rm and} \quad u_{N - k}: = e^{i \,{\rm arg}(b_k)}\arctan{|\frac{b_k}{a_k}|}.
\]
\item Define ${\cal A}_{k + 1}$ by
\[
{\cal A}_{k + 1}(z) = E( - y_k, z)\cdot R^{- 1}(u_{N - k})
\cdot E( y_k, z) \cdot {\cal A}_k(z),
\]
where matrices $E$ and $R$ are given by (\ref{E-limit}) and (\ref{R-limit}), respectively.
\end{enumerate}

We repeat the sequence of operations (i) to (iv) until we reach ${\cal A}_{N - 1}$. Then the
values, obtained in (iii), provide the wanted array
\[
{\cal U} = \{ (x_1, u_1), (x_2, u_2), \ldots (x_N, u_N) \}.
\]
\label{thmD}
\end{thm}
\epf

\noindent In the proof of our theorem we will need the following lemma:

\begin{lemma}
For every $k = 0, \ldots, \lceil \frac{N}{2} \rceil$ and for every ordered array
$n_1, n_2, \ldots n_{2k - 1}$
of odd number of elements from $\{n_j\}$
we have
\[
x_N \ > \ x_{n_{2k - 1}} - x_{n_{2k - 2}} + x_{n_{2 k - 3}} + \ldots - x_{n_2 } + x_{n_1} .
\]
\label{lemmarightmost}
\end{lemma}

\noindent {\bf Proof:}
Consider the real numbers
\[
x_{n_{2k - 1}} - x_{n_{2k - 2}} + x_{n_{2 k - 3}} - \ldots - x_{n_2 } + x_{n_1}
\]
Because of the orderings of
$\{n_j\}_{j = 1, \ldots, N}$ and
$\{x_j \}_{j = 1, \ldots , N}$, we have
for every monotonically increasing array $n_1, n_2, \ldots n_{2k - 1}$ of integers,
\[
x_{n_{2 k - 1}} \ > \ x_{n_{2k - 1}} - (x_{n_{2 k - 2}} - x_{n_{2k - 3}}) - \ldots -
(x_{n_2} - x_{n_1}).
\]
Indeed, on the right, the positive numbers $x_{n_{j +1}} - x_{n_j}$ are
subtracted from $x_{n_{2 k - 1}}$.
The number $x_N$ is the largest among all $x_n$, so the lemma is proved.

\epf

\noindent{\bf  Proof of Theorem \ref{thmD}:}
Our task is to find
the array of pairs of numbers 
\[
{\cal U} = \{(x_1, u_1), (x_2, u_2), \ldots, (x_N, u_N)\}
\]
which determine the distribution $u$.
Let us write formula (\ref{thirddelta}) more explicitly. 
First, we introduce the function ${\cal S}(u)$ of 
$\{u_n\}_{n = 1, \ldots, N} = \{ e^{i \phi_n} r_n \}_{n = 1, \ldots, N}$, using the formula
\[ 
{\cal S}_{n_1, n_2, \ldots , n_d}(u) = 
\prod_{k = 1}^d e^{(- 1)^{d - k} \phi_{n_k}}\sin{(r_{n_k})} \! \! \! \! \! \!
\prod _{m \ne n_k; \, k = 1, \ldots, d} \! \! \! \cos{( r_m)}.
\]
\label{functionsSC}
Let ${\cal A}(z) = {\cal F}_r^D[u](z)$ for some 
$u = \sum_{n = 1}^N u_n \, \delta_{x_n}$. We have
\[
{\cal A}(z) = {\cal F}_r^D[u](z) = \pmatrix{ {\cal C}(u) + A[u](z) & B[u](z) \cr
- \overline{B[u](z)} & {\cal C}(u) - \overline{A[u](z)} },
\]
where
\begin{eqnarray} \fl
A[u](z) & =& \sum_{k = 1}^{\lfloor \frac{N}{2} \rfloor} (- 1)^{k},
\sum_{ n_{2 k} > n_{2 k - 1} > \ldots > n_1 }
e^{- 2 \pi i \, (x_{n_{2 k}} - x_{n_{2k - 1}} + \ldots + x_{n_2} - x_{n_1})\cdot z}
{\cal S}_{n_1, n_2, \ldots , n_{2k}}(u)
\label{evenpart}
\\ \fl
B[u](z) & = & \sum_{k = 1}^{\lceil \frac{N}{2} \rceil} (- 1)^{k + 1} \! \! \! \! \!
\sum_{ n_{2 k - 1} > n_{2 k} > \ldots > n_1 }
e^{- 2 \pi i (x_{n_{2k -1}} - x_{n_{2 k} }+ \ldots - x_{n_2}+ x_{n_1}) \cdot z}
{\cal S}_{n_1, n_2, \ldots , n_{2k - 1}}(u).
\label{oddpart}
\end{eqnarray}

Let $\widehat{w}(z) = \sum _{ j = 1}^N w_j \, e^{ -2 \pi i x_j z}$. Recall that the inverse linear Fourier transform of $\widehat{w}$ is given by
\[
F^{- 1} [\widehat{w}](x) = \sum_{j = 1}^N w_j \, \delta _{x_j}(x).
\]
We can apply this formula to the function $B[u](z)$, given by (\ref{oddpart}), and obtain
\[ \fl
F^{- 1}[B[u]]= {\cal S}_N (u)\, \delta _{x_N} + \sum_n^{N - 1} {\cal S}_n (u)\,
\delta _{x_n}
+
\sum_{k = 2}^{\lceil \frac{N}{2} \rceil} (- 1)^k \! \! \! \! \!
\sum_{{\cal N}(2k - 1)}
\! \! \! \!\! {\cal S}_{n_1, n_2, \ldots , n_{2k - 1}}(u) \ \delta_{x_{n_1, \ldots, n_{2k - 1}}},
\]
where ${\cal N}(2k - 1)$ is the set of all increasing integer arrays
$0< n_1 < n_2 < \ldots < n_{2k - 1} \leq N$ and
$x_{n_1, \ldots, n_{2k - 1}} = x_{n_{2k - 1}} - x_{n_{2k - 2}} + \ldots - x_{n_2 }+ x_{n_1}$.
Therefore, for a generic choice of $\{x_n\}_{n = 1, \ldots, N}$,
the linear inverse Fourier transform of $B[u]$ is the sum of the $ 2^{N - 1}$ delta functions.
Lemma \ref{lemmarightmost} tells us that ${\cal S}_N(u) \, \delta_{x_N}$ is
precisely the rightmost of these delta functions.

The linear inverse Fourier transform of the upper-left term of ${\cal F}^D[u]$ is given by
\[ \fl
F^{- 1}[{\cal C}(u) + A[u](z)] = {\cal C}(u) \, \delta_0 \ +
\sum_{k = 1}^{\lfloor \frac{N}{2} \rfloor} (- 1)^{k}
\sum_{ {\cal N}(2 k) }
{\cal S}_{n_1, n_2, \ldots , n_{2k}}(u) \ \delta_{x_{n_1, \ldots, n_{2k}}},
\]
where ${\cal N}(2 k )$ is the set of all increasing integer arrays.
$0< n_1 < n_2 < \ldots < n_{2k} \leq N$, and
$x_{n_1, \ldots, n_{2k}} = x_{n_{2k}} - x_{n_{2k - 1}} + \ldots + x_{n_2 }- x_{n_1}$.
In the above sum, ${\cal C}(u) \delta _0$ is the only delta term, located at $x = 0 $.

From the linear inverse transforms $F^{ - 1}[B[u]]$ and $F^{- 1}[{\cal C}(u) + A[u]]$, we can read
the values $x_N$, ${\cal S}_N(u)$, and ${\cal C}(u)$. \label{read-off}
From the latter two we get
\label{DValues}
\[
\frac{{\cal S}_N(u)}{{\cal C}(u)} =
\frac{e^{i \phi _N} \sin{( r_N})\prod_{n = 1}^{N - 1} \cos{( r_n)}}{\prod_{n = 1}^{N} \cos{(r_n)}}
= e^{i \phi_N} \tan{( r_N)}
\]
which gives
\[
u_N = e^{i \phi_N} \arctan{|\frac{{\cal S}_N(u)}{{\cal C}(u)}|}.
\]
We have obtained our first pair $(x_N, u_N)$ of the array
$
{\cal U} = \{ (x_n, u_n); n = 1, \ldots N\}.
$
To find another pair of ${\cal U}$, we can proceed as follows: Recall the formula (\ref{FDAd})
\[ 
{\cal F}_r^D[u](z) = \prod_{n = N}^1 \Ad_{E(- x_n, z)} R(u_n).
\]
Thus, the matrix function
\begin{eqnarray*} \fl
{\cal A}_1 & = & E(- x_N, z) \cdot R(u_N)^{- 1} \cdot E( x_N, z) \cdot 
{\cal A}\\ \fl
& = & E( - x_{N - 1}, z) \cdot R(u_{N - 1}) \cdot E( x_{N - 1}, z)\ \cdots \
E( - x_1, z) \cdot \ R( u_1) \cdot E(x_1, z) \\ \fl
& = & \prod_{n = N - 1}^1 \Ad_{E(- x_n, z)} R(u_n) 
\end{eqnarray*}
is the nonlinear Fourier transform ${\cal F}_r^D[u^{(1)}]$ of the distribution
\[
u^{(1)} = \sum_{n = 1}^{N - 1} u_n \, \delta_{x_n}.
\]
Now we can repeat the above procedure with ${\cal A}_1 = {\cal F}_r^D[u^{(1)}]$ instead of
${\cal A}(z)$, and obtain the next pair
$(x_{N - 1}, u_{N - 1})$ of the array ${\cal U}$.
Repeating this procedure $N$ times yields all the pairs in ${\cal U}$. Hence our theorem is proved.

\epf

\noindent An immediate consequence of the theorem is the following:
\begin{cor}
Let $z \longmapsto {\cal A}(z)$ be an $SU(2)$-valued function. Then, ${\cal A}(z)$ is the nonlinear Fourier
transform ${\cal F}_r^D[u](z)$ for some distribution
$u = \sum _{n = 1}^N u_n \, \delta_{x_n}$
if and only if ${\cal A}_{N} = I$. 
\end{cor}

\begin{rmk}
Suppose $\widehat{\cal A} = {\cal F}_r^D[\widehat{u}](z)$, where 
$\widehat{u} = \sum_{n = 1}^N \Delta x_n u_n \, \delta_{x_n}$. Then, by applying the above equation,
algorithm yields array $\widehat{\cal U} = \{ (x_1, \Delta x_1 u_1), 
(x_2, \Delta x_2 u_2), \ldots (x_N, \Delta x_N u_N) \}$. We obtain the array ${\cal U}$ by
dividing the second components of $\widehat{\cal U}$ by $\Delta x_n$ which we obtain by subtracting 
the appropriate first components; $\Delta x_n = x_{n + 1} - x_n$.
\end{rmk}

\section{The inverse of ${\cal F}^E$}

In this section, $z$ takes values in the set $\{0, 1, \ldots, N - 1\}$.
Let us denote
\begin{equation}
{\cal A}(z)=  \pmatrix{ \alpha^E(z) & \beta^E(z) \cr
- \overline{\alpha ^E(z)} & \overline{\beta^E(z)} }     \quad z = 0, \ldots, N - 1 
\label{QSU}
\end{equation}
and let us recall the definitions (\ref{EandJ}) of the matrices
 $E_N(n, z)$ and $U_n$.

 We shall prove the following theorem:

\begin{thm}
Let ${\cal A}(z)$, $z = 0, 1, \ldots, N - 1$ be an $N$-tuple of matrices of the form (\ref{QSU}). 
Suppose we know that there exists a vector 
$u = (u_0, u_1, \ldots , u_{N - 1})$ such that
\[
{\cal A}(z) = {\cal F}^E[u](z).
\]
Then $u$ can be computed by the following algorithm.

Set ${\cal A}_0(z) = {\cal A}(z)$.

The matrix ${\cal A}_{k + 1}(z)$ is obtained from ${\cal A}_k(z)$, where
\[
{\cal A}_k(z) = \pmatrix{ a_k(z) & b_k(z) \cr
- \overline{b_k(z)} & \overline{a_k(z)} }
\]
by the following steps.
\begin{enumerate}
\item Compute the inverse linear discrete Fourier transform $(F^D)^{- 1}[b_k](n)$.
\item Set
\[ 
\frac{u_{N - k}}{N} = \Bigl((F^D)^{ - 1}[b_k]\Bigr) (N - 1),
\]
and
\[
\widehat{R_k}(z) = ( I + \pmatrix{ 0 & e^{ - 2 \pi i \frac{(N - 1) z}{N}} \ \frac{u_{N - k}}{N}\cr
- e^{2 \pi i \frac{(N - 1) z}{N}} \ \frac{\overline{u_{N - k}}}{N} & 0 } ).
\]

\item Define
\[ \fl
{\cal A}_{k + 1}(z) = \Ad_{E_N(- 2, z)} \Bigr[ \widehat{R_k}(z)^{- 1} \cdot {\cal A}_k (z)\Bigl] 
\cdot (I + \frac{1}{N} U_{N - k})
\]
for every $z = 0, 1, \ldots , N - 1$.
\end{enumerate}
Then $u = (u_0, u_1, \ldots , u_{N - 1})$, where 
$\frac{u_n}{N}$ are given by (ii), is the solution
of the equation
\[
{\cal F}^E[u](z) = {\cal A}(z).
\]
\label{Ethm}
\end{thm}

\noindent  {\bf Proof:}
Recall that the discrete Fourier transform ${\cal F}^E[u](z)$ of a discrete function
$u = (u_0, u_1, \ldots , u_{N - 1})$ can be given by its Dyson form 
\begin{equation} \fl
{\cal F}^E[u](z) = I + \sum_{d = 1}^N \Bigl( \frac{1}{N^d}
\sum_{\Delta_d^D}
L^G(n_d, z) \cdots L^G(n_2, z) \cdot L^G(n_1, z) \Bigr),
\label{EulerF11}
\end{equation}
where $\Delta_d^D$ denotes the discrete ordered simplex
\[
\Delta_d^D = \{ (n_1, n_2, \ldots, n_d) \in \mathbb{Z}^d; \ 0 \leq n_1 < n_2 < \ldots < n_d \leq N- 1\}.
\]
We have rewritten ${\cal F}^E[u]$ more explicitly as

\begin{eqnarray} \fl \nonumber
{\cal F}^E[u](z) = I & +& \sum_{n = 1}^N E_N(- 2n, z) \, \frac{1}{N} \, U_n \\ 
& + & 
\sum_{d = 2}^N \ \ \sum_{ N - 1\geq n_d > \ldots, n_2> n_1 \geq 0} \! (\frac{1}{N})^d
E_N \Bigl(- 2 \sum_{k = d}^1 (- 1)^{d - k} n_k , z\Bigr) \, \prod _{k = d}^1 U_{n_k}
\label{EulerF22}.
\end{eqnarray}
We can express ${\cal F}^E[u]$ as the linear discrete Fourier transform of the nonlinear function of $u$
in the following way. Let us stratify $\Delta_d^D$ into subsets $D_d(l)$, defined by
\[ \fl
D_d(l) = \{(n_1, n_2, \ldots, n_d) \in \Delta _d^D; \ n_d - n_{d - 1} + n_{d - 2} + 
\ldots + (- 1)^{d - 1} n_1 = l \}.
\] 
Above, $l \in \{0, 1, \ldots , N - 1\}$. (For $d > 2$, sets $D_d(0)$ and $D_d(N - 1)$ are empty.)
Now, we can write (\ref{EulerF22}) in the form
\begin{equation}
\fl
{\cal F}^E[u](z) = I + \sum _{l = 0}^{N - 1} E_N(- 2 l, z)  \cdot  
\sum_{d = 1}^N \Bigl( (\frac{1}{N})^d \! \! \! \! \sum_{(n_1, \ldots , n_d) \in D_d(l)} 
\ \prod_{k = d}^1 U_{n_k} \ \Bigr) .
\label{EpolytopeF}
\end{equation}
\label{reorganize}
It is easily seen from (\ref{EulerTransfProd}) 
that for every $z$, the transform ${\cal F}^E[u](z)$ is a matrix of the form
(\ref{QSU}).
(In general, the determinant of ${\cal F}^E[u](z)$ is not equal to $1$ and is therefore not an element
of $SU(2)$.)
The entries $\alpha^E(z)$ and $\beta^E(z)$ which determine the matrix are given by
\begin{eqnarray} \fl
\alpha^E(z) & = & 1 + \sum_{l = 0}^{N - 1} e^{- 2 \pi i \frac{l z}{N}} \
\Bigl( \ \sum_{ k = 1}^{\lfloor \frac{N}{2}\rfloor} (- 1)^k \ (\frac{1}{N})^{2 k} \ 
\! \! \! \! \sum_{(n_{1}, \ldots , n_{2 k}) \in D_{2 k }(l)} 
\ u_{n_{2k}} \overline{u_{n_{2k - 1}}} \cdots u_{n_2} \overline{u_{n_{1}}} \Bigr) \\ \fl \nonumber
& & \\ \fl
\beta^E(z) & = & \sum_{l = 0}^{N - 1} e^{- 2 \pi i \frac{l z}{N}} \ 
\Bigl(\ \sum_{ k = 1}^{\lceil \frac{N}{2}\rceil} (- 1)^{k - 1} \ (\frac{1}{N})^{2 k - 1} \ 
\! \! \! \! \! \! \! \! \! \sum_{(n_1, \ldots , n_{2 k - 1}) \in D_{2 k -1 }(l)} 
u_{n_{2k - 1}} \overline{u_{n_{2 k - 2}}} \cdots \overline{u_{n_2}}u_{n_1} \Bigr).
\label{betaterm}
\end{eqnarray}
Note that in the above equations the expressions inside the large brackets, next
to the factor $e^{ - 2 \pi i \frac{l z}{N}}$ are functions of variable $l$, because they
are determined by  the sets $D_n(l)$
Note also that the $k = 1$ term in (\ref{betaterm}) is the usual linear discrete 
Fourier transform $F^D[u](z)$ of $u$. 
Let us denote
\[
\Sa (l, k) = \sum_{ k = 1}^{\lceil \frac{N}{2}\rceil} (- 1)^{k - 1} \ (\frac{1}{N})^{2 k - 1} \ 
\! \! \! \! \sum_{(n_1, \ldots , n_{2 k - 1}) \in D_{2 k -1 }(l)} 
u_{n_{2k - 1}} \overline{u_{n_{2 k - 2}}} \cdots \overline{u_{n_2}} \, u_{n_1}.
\]
\label{Ealgorithm}
Then 
\[
\beta ^E(z) = \sum_{l = 0}^{N - 1} e^{ - 2 \pi \frac{l z}{N}} \ 
\  \Sa(l, k)  =
F^D \Bigl[  \Sa(l, k) \Bigr] (z),
\]
Therefore, applying the
linear inverse discrete transform to both sides of the above equation gives
\begin{equation}
\Sa(l, k) = 
\Bigl((F^D)^{- 1}[\beta^E] \Bigr) (l), \quad l = 0, \ldots, N - 1.
\label{invsystem}
\end{equation}
The right-hand sides of the above equations are known. Therefore, we have $N$ equations for the
$N$ unknowns $u_0, u_1, \ldots , u_{N - 1}$. The above system is highly nonlinear. The terms
$\Sa(l, k)$ are polynomials of degrees $2 k - 1$ in the variables $u_0, u_1, \ldots, u_{N - 1}$. The 
number of terms of  degree $2k - 1$ of such polynomial is equal to
the number of points in the stratum $D_{2 k- 1}(l)$.
Recall that the number of points in $D_{2k - 1}(l)$ is the number 
of solutions of the equation 
\[ 
n_{2k - 1} - n_{2k - 2} + n_{2k - 3} - \ldots - n_2 + n_1 = l, 
\]
where $ N- 1\geq n_{2k - 1} > n_{2k - 2} > \ldots > n_1 \geq 0$
are integers. For $l = N - 1$, the above equation has only one solution:
$n_1 = N - 1$. Therefore, the last equation of system (\ref{invsystem}) is:
\[
\frac{u_{N - 1}}{N} = \Bigl( F^{- 1}[\beta^E] \Bigr) (N - 1).
\]
We have found the last component $u_{N- 1}$ of the unknown discrete function 
$(u_0, u_1, \ldots, u_{N - 1})$.

Recall that ${\cal F}^E[u]$ is given by (\ref{EulerTransfProd}); that is,
\[ \fl
{\cal F}^E[u](z) = \Bigl(I + \frac{1}{N} L^G(N - 1, z)\Bigr
) \cdot \Bigl(I + \frac{1}{N} L^G(N - 2, z)\Bigr) 
\cdots \Bigl(I + \frac{1}{N} L^G(0, z)\Bigr).
\]
Let for every $z = 0, 1, \ldots , N - 1$,
\[ \fl
{\cal A}_1(z) = \Ad_{E_N(- 2, z)}\Bigl[
\Bigl(I + \frac{1}{N} L^G(N - 1, z)\Bigr)^{- 1} \cdot {\cal F}^E[u](z) \Bigr]
\cdot (I + \frac{1}{N} U_{N - 1})
\]
A rather straightforward calculation shows that
\[
{\cal A}_1(z) = {\cal F}^E[u^{(1)}](z),
\]
where $u^{(1)}= (u_{N - 1}, u_0, u_1, \ldots, u_{N - 2})$
is $u$ with components cyclically permuted by one step.

Now we repeat the procedure with ${\cal A}_1$ in place of ${\cal A}$: We calculate the
linear  discrete inverse Fourier transform $(F^{D})^{- 1}[b_1]$, where
\[
{\cal A}_1(z) = \pmatrix{ a_1(z) & b_1(z) \cr
                                        - \overline{b_1(z)} & \overline{a_1(z)} },
                                        \]
solve the equation
\[
\frac{u_{N - 2}}{N} = \Bigl( (F^{D})^{ - 1}[b_1]\Bigr)(N - 1)
\]
and set
\[ \fl
{\cal A}_{2}(z) = \Ad_{E_N(- 2, z)} \Bigr[ \widehat{R_1}(z)^{- 1} \cdot {\cal A}_1 (z)\Bigl] 
\cdot (I + \frac{1}{N} U_{N - 2}),
\]
where
\[
\widehat{R_1}(z) = ( I + \pmatrix{ 0 & e^{ - 2 \pi i \frac{(N - 1) z}{N}} \ \frac{u_{N - 2}}{N}\cr
- e^{2 \pi i \frac{(N - 1) z}{N}} \ \frac{\overline{u_{N - 2}}}{N} & 0 } ).
\]
Repeating this procedure $N$  times yields all the wanted values $(u_0, u_1, \ldots , u_{N - 1})$.

\epf

\begin{cor}
A matrix function
$
z \longmapsto {\cal A}(z); \quad z = 0, 1, \ldots, N - 1,
$
 of form (\ref{QSU}),
is equal to the discrete nonlinear Fourier transform ${\cal F}^E[u](z)$ for some $u$  
if and only if
\[
{\cal A}_{N}(z) = {\cal A}(z)
\]
for every $z$.
\end{cor}

\begin{rmk}
The starting point of the algorithm of Theorem \ref{Ethm} is the application of
the inverse linear discrete Fourier transform to ${\cal A}(z)$. This yields the nonlinear 
system
\[\fl
\sum_{k = 1}^{\lceil\frac{N}{2}\rceil} (- 1)^{k - 1} \ (\frac{1}{N})^{2k - 1} \! \! \! \! \! \! \!
\sum_{(n_1, \ldots , n_{2 k - 1}) \in D_{2 k -1 }(l)} 
u_{n_{2 k - 1}} \overline{u_{n_{2k - 2}}} \cdots \overline{u_1} = 
(F^D)^{- 1}(\beta^E](l)
\]
for $l = 0, \ldots, N - 1$ and 
for the unknowns $u_0, u_1, \ldots u_{N - 1}$. The left-hand sides for $l$ close to $\frac{N}{2}$
are huge. More precisely, the number of elements in each $D_{2k - 1}(l)$ is equal to
\[
\sharp D_{2 k - 1}(l) = {l \choose k - 1} {N - l - 1 \choose k - 1}.
\]
For $N = 100$, $k = 10$ and $l = 25$ we have $\sharp D_{2 k - 1}(l) \sim 2,3 \times 10^{17}$.
It turns out that for fixed $N$ and $k$, the numbers $\sharp D_{2 k - 1}(l)$
are essentially (after normalization) distributed according to a discretisation 
of {\rm Beta } distribution. For more details on this, see \cite{SaProb}.
\end{rmk}

\section{Dual transforms ${\cal F}^D$, ${\cal F}^D_d$ and the constant mass distributions}
\label{dualitysec}

Throughout this section, we shall restrict ourselves to distributions
$u = \sum _{n = 1}^N u_n \, \delta_{x_n}$ with real positive coefficients, $u_n > 0$, and
$\sum_{n = 1}^N u_n < 1$. As before, let $0 < x_1 < x_2 < \ldots < x_n < 1$.
The central point will be the fact that the roles of variables $u_n$ and
$\Delta x_n$ in ${\cal F}^D$ are, to an extent, symmetrical. For easier reading, we denote:
$\xi_n = \Delta x_n$ in this section.

Consider two
discrete functions defined on $\{0, 1, \ldots, N + 1\}$ by
\begin{equation}
\fl
x : n \mapsto x_n = \sum_{k = 0}^{n - 1} \xi_k, \ \ \ x_{N + 1} = 1 \quad \quad {\rm and } \quad
\quad
v : n \mapsto v_n = \sum_{k = 0}^n u_k, \ \ \ v_{N + 1} = 1.
\label{symmfns}
\end{equation}
Note that
\[
\xi _n = x_{n + 1} - x_n \quad {\rm and} \quad u_n = v_n - v_{n - 1}.
\]
We can define two transforms of the pair of functions $(x, v)$ by
\begin{eqnarray*}
{\cal G}[x, v](z) & = & \Bigl( \prod_{n = N + 1}^1 R(v_n - v_{n - 1}) \cdot
E(x_n - x_{n - 1}, z) \Bigr) \cdot R(v_0) \\
{\cal G}_d[x, v](z) & = & \Bigl( \prod_{n = N + 1}^1 R(v_n - v_{n - 1}, z) \cdot
E(x_n - x_{n - 1}) \Bigr) \cdot R(v_0),
\end{eqnarray*}
where
\[ 
R(v, z) = \pmatrix{ \cos{(\pi v z)} & \sin{(\pi v z)} \cr
- \sin{(\pi v z)} & \cos{(\pi v z)} }, \quad \quad
E(x, z) = \pmatrix{ e^{\pi i x z} & 0 \cr
0 & e^{- \pi i x z} },
\]
and $R(v) = R(v, \frac{1}{\pi z})$ and $E(x) = E(x, \frac{1}{ \pi z}) = {\rm diag} (e^{i x}, e^{- i x})$.

Let  $v$ be an arbitrary function from (\ref{symmfns}) and let 
$v_r$ be its restriction on $\{1, 2, \ldots, N \}$. Then we denote
\[\fl 
R(1 - v_{N })^{- 1} \cdot {\cal G}[x, v] \cdot R(v_0)^{- 1} = 
{\cal G}[x, v_r](z) = {\cal F}^D[\xi, u](z) ,\quad {\rm for} \quad 
u = \sum_{n = 1}^N u_n \, \delta _{x_n}.
\]
Note that
\[
{\cal F}^D[\xi, u](z) = {\cal F}^D[u](z),
\]
where ${\cal F}^D[u](z)$ is the transform studied above and introduced in Definition \ref{FD}.
\begin{defn} The transform ${\cal G}_d[x, v](z)$ is called the dual nonlinear Fourier transform, associated to ${\cal F}^D[\xi, u](z)$. In variables $(\xi, u)$ 
we shall denote it by ${\cal F}^D_d[\xi, u](z)$.
\end{defn}
Note that to obtain ${\cal F}_d^D$ from ${\cal F}^D$, we have to choose a value $u_0$.
The value $u_{N + 1}$ is then determined by the condition $v(N+ 1) = 1$.
A more explicit expression of the dual transform is
\[
{\cal F}^D_d[\xi, u](z) = R(1, z) \cdot \prod _{n = N}^0 \Ad _{R( - v_n, z)} E(\xi_n)
\]
which is analogous to the expression (\ref{AdRep1}).
Let now
\[
A = \pmatrix{ i &- i \cr
1 & 1},
\]
and denote $\widetilde{{\cal F}^D_d}[\xi, u](z) = A^{-1} \cdot {\cal F}^D_d[\xi, u] \cdot A$.
We get
\[
\widetilde{{\cal F}^D_d}[\xi, u](\zeta) = E(1, \zeta)
\cdot \prod_{n = N}^0 \Ad_{E(- v_n, \zeta)} \widehat{R}(\xi_n),
\]
where $ \zeta = - z$ and
\begin{equation}
\widehat{R}(\xi_n) = \pmatrix{ \cos{ \xi_n} & - i \sin{\xi_n} \cr
- i \sin{\xi_n} & \cos{\xi_n} } = \cos{\xi_n} \, I + \sin{\xi_n} \, L.
\label{Rmathat}
\end{equation}
Let us introduce the reduced form $\widehat{{\cal F}^D_d}[\xi, u](\zeta)$ of 
$\widetilde{{\cal F}^D_d}[\xi, u](\zeta)$ by
\begin{equation}\fl
\widehat{{\cal F}^D_d}[\xi, u](\zeta) =  \prod_{n = N}^0 \Ad_{E(- v_n, \zeta)} \widehat{R}(\xi_n) = 
\prod_{n = N }^0 \pmatrix{ \cos{\xi_n} & - e^{- 2 \pi i v_n \zeta} \ i \sin{\xi _n} \cr
- e^{2 \pi i v_n \zeta} \ i \sin{\xi_n} & \cos{\xi_n} }.
\label{dualexp}
\end{equation}      
Formulae (\ref{dualexp}) and  (\ref{FDAd}) for the reduced ${\cal F}^D_r$ are analogous. 
We only have to replace
$v_n$ by $x_n$ and $\xi_n = \Delta x_n$ by $u_n$ to get
one from the other.

Under the assumptions on $\{u_n\}$ listed at the beginning of this section, the use
of ${\cal F}^D$ or ${\cal F}^D_d$ in the study of distributions
$\sum_{n = 0}^N u_n \, \delta_{x_n}$ yields essentially the same results. However, if one
of the two variables $\{\xi_n\}$ or $\{u_n\}$ is constant, then the correct choice of one of the two transforms can considerably reduce the amount of work.

\paragraph{\it \underline{Shorter algorithm for searching distributions of the form
$w = \sum_{n = 1}^N u_c \, \delta_{x_n}$}:}
Let ${\cal A}(z)$ be equal to ${\cal F}^D[u](z)$ for some $u$ of the form:
$u = \sum_{n = 1}^N u_n \, \delta_{\frac{n}{N}}$.
One can expect that the inverse
transform $({\cal F}^D)^{- 1}$ of ${\cal A}(z)$ can be computed using the algorithm
for $({\cal F}^E)^{- 1}$ which is much faster than that for $({\cal F}^D)^{- 1}$.
Analogously,
suppose ${\cal B}(\zeta)$ is the $\widehat{{\cal F}^D_d}$ - transform 
of distribution of the form
$
w = \sum_{ n= 1}^N u_c \, \delta_{x_n}
$,
where $u_c $ is a constant. Denote $M = N + 1$ and let, for convenience, 
$u_c = \frac{1}{M}$. Then we have
the following theorem:
\begin{thm}
Let
\[
{\cal B}(\zeta) = \widehat{{\cal F}^D_d}[w](\zeta)  =  \widehat{{\cal F}^D_d}[\xi, u_c](\zeta).
\]
Then the array $\{x_1, x_2, \ldots , x_N\}$ which determines 
$w = \sum_{n = 1}^N u_c \, \delta_{x_n}$ can be found using the algorithm for the calculation
of $({\cal F}^E)^{- 1}$, described in Theorem \ref{Ethm}.
\label{dual}
\end{thm}

\noindent{\bf Proof:} 
We introduce
the functions, analogous to those on page
\pageref{functionsSC}:
\[ \fl
{\cal S}_{n_1, n_2, \ldots , n_d}(\xi) = \prod_{k = 1}^d \sin{\xi_{n_k}} \! \! \!
\prod _{m \ne n_k; \, k = 0, \ldots, d} \cos{\xi_m}, \quad \ \ {\rm and} \quad \ \
{\cal C}(\xi) = \prod_{n = 0}^N \cos{\xi_n}.
\]
Again, we have the commutation relation
\[
E(v, \zeta) \cdot L = L \cdot E(- v, \zeta), \quad \quad L = \pmatrix{ 0 & - i \cr
- i & 0 }
\]
where $L$ appears in (\ref{Rmathat}).
This enables us to write
\begin{eqnarray*} \fl \nonumber
\widehat{{\cal F}^D_d}[w](\zeta) = {\cal C}(\xi) \, I \, & + &
\sum_{n = 0}^{M - 1} E(- 2 \frac{n}{M}, \zeta) \cdot {\cal S}_n(\xi)\cdot L \\
\fl
& + & \sum_{d = 2}^{M - 1} \sum_{M > n_d > \ldots > n_1 \geq 0} \! \! \! \! \!
E(-2 \sum_{k = 1}^d (- 1)^{d - k} \frac{n_k}{M}, \zeta)
\cdot {\cal S}_{n_1, n_2, \ldots , n_d} (\xi) \cdot L^d.
\end{eqnarray*}
Now, we evaluate 
${\cal C}(\xi)$, as on page \pageref{DValues} in Section \ref{InvD}. (Here we have to apply the ``full''
inverse linear Fourier transform, but we have to do it only once and not $2N$ times as in 
Section \ref{InvD}.)
If we divide the two sides of the above equation by ${\cal C}(\xi)$,
we get
\[ \fl
{\cal H}[w](\zeta) =
I + \sum _{l = 0}^{M - 1} E(- 2 \frac{l}{M}, \zeta) \ \cdot \
\sum_{d = 1}^{M - 1} \Bigl( \sum_{(n_1, \ldots , n_d) \in D_d(l)}
\ \prod_{k = 1}^d \tan{(\xi_{n_k}}) \ \ \Bigr) \cdot \ L^d \Bigr).
\label{dualpolytopetransfA}
\]
As before,
\[ \fl
D_d(l) = \{(n_1, n_2, \ldots, n_d) \in \Delta _d^D; \ n_d - n_{d - 1} + n_{d - 2} +
\ldots + (- 1)^{d - 1} n_1 = l \}.
\]
From formula (\ref{EpolytopeF}) and discussion leading to it, we see that
${\cal H}[w](\zeta)$ is the Euler type discrete nonlinear transform given by 
initial value problem
\[
\Psi(n + 1, \zeta) - \Psi(n, \zeta) = K^G (n, \zeta) \cdot \Psi(n, \zeta), \quad \Psi(0, \zeta) = I,
\]
where the coefficient matrix is 
\[
K^G(N, \zeta) = \pmatrix{ 0 & - e^{- \frac{2 \pi i n \zeta}{M}} i \, \tan(\xi_n) \cr
- e^{\frac{2 \pi i n \zeta}{M}} i \, \tan(\xi_n) & 0 }.
\]
It is now clear that, with suitable modifications, we can apply the algorithm from the  Theorem
\ref{Ethm} for the evaluation of $\tan{(\xi_n)}$, $n =0, \ldots, N$.

The terms $\alpha^D(\zeta)$ and $\beta ^D(\zeta)$ of the matrix
\[
{\cal H}[w](\zeta) = \pmatrix{\alpha^D(\zeta) & \beta^D(\zeta) \cr
- \overline{\beta^D(\zeta)} & \overline{\alpha^D(\zeta)} }
\]
are given by
\begin{eqnarray} \fl
\alpha^D(\zeta) & = & 1 + \sum_{l = 0}^{M - 1} e^{- 2 \pi i \frac{l \zeta}{M}} \
\Bigl( \ \sum_{ k = 1}^{\lfloor \frac{M}{2}\rfloor} (- 1)^k \
\! \! \! \! \sum_{(n_1, \ldots , n_{2 k}) \in D_{2 k }(l)} \! \! \! \! \! \!
\tan{\xi_{n_1}}\tan{\xi_{n_2}} \cdots \tan{\xi_{n_{2 k}}} \Bigr) \\ \fl \nonumber
& & \\ \fl
\beta^D(\zeta) & = & \sum_{l = 0}^{M - 1} e^{- 2 \pi i \frac{l \zeta}{M}} \
\Bigl(\ \sum_{ k = 1}^{\lceil \frac{M}{2}\rceil} i (- 1)^k \
\! \! \! \! \! \! \! \! \! \! \sum_{(n_1, \ldots , n_{2 k - 1}) \in D_{2 k -1 }(l)}
\! \! \! \! \! \! \! \! \! \! \! \!
\tan{\xi_{n_1}}\tan{\xi_{n_2}} \cdots \tan{\xi_{n_{2 k - 1}}} \Bigr).
\label{betadAA}
\end{eqnarray}
Denote
\[
\Sa _r(l, k) = \sum_{ k = 1}^{\lceil \frac{M}{2}\rceil} i (- 1)^k
\sum_{(n_1, \ldots , n_{2 k - 1}) \in D_{2 k -1 }(l)}
\! \! \! \! \! \! \! \! \! \! \! \!
\tan{\xi_{n_1}}\tan{\xi_{n_2}} \cdots \tan{\xi_{n_{2 k - 1}}}.
\]
Similar to page \pageref{Ealgorithm}, we obtain from (\ref{betadAA}):
\[
 \Sa_r(l, k) = (F^D)^{- 1} [ - i \beta^D](l)
\]
and, taking into account that $M - 1 = N$, we have
\[
\tan{\xi_{N}} = (F^D)^{- 1}[- i \beta^D](N).
\]
Here $F^D$ denotes the linear discrete Fourier transform.
From here on, we follow the algorithm of Theorem \ref{Ethm} without changes
to obtain all $\xi_n$, and $x_n = \sum_{k = 0}^{n - 1}\xi_k$. 

\epf

\section{Conclusions} We described two  algorithms for evaluating
the inverses of the discretisation ${\cal F}^E$ and ${\cal F}^D$ of the nonlinear Fourier
transform ${\cal F}$. The two algorithms have similar structures. In both cases, we apply
the inverse {\it linear} Fourier transform on
the given matrix function ${\cal A}(z)$, where $z$ is discrete in the case of ${\cal F}^E$ and 
continuous in the case of ${\cal F}^D$. This yields
highly nonlinear systems for the unknown discrete function $\{ u_n\}_{n = 0, \ldots, N - 1}$,
and also for $\{x_n\}_{n = 0, \ldots N- 1}$ in the case of ${\cal F}^D$.
However, in both cases, the ``last'' equation in these systems is trivial and yields the
unknowns $u_{N - 1}$ or $(x_{N - 1}, u_{N - 1})$. The structures of ${\cal F}^E$ and ${\cal F}^D$ allow us to
find the other unknowns recursively.

Our primary interest in these algorithms is their mathematical content and not numerical efficiency.
We nevertheless comment shortly on the computational complexity of the
algorithms.
Of these two, the case of ${\cal F}^D$ is numerically more demanding.
We have to calculate good approximations of continuous linear Fourier transforms $N$ times.
These would, of course, be calculated using FFT, but the data size of these FFTs 
must be considerably larger than $N$. 

The inverse of ${\cal F}^E$ can be solved by $N$ evaluations of the FFT with data size $N$:
so the order of complexity is $N^2 \log{N}$. Suppose we have
a situation in which ${\cal F}^E[u](z)$
is known not only for the values $z \in \{\frac{n}{N}\}_{n= 0, \ldots, N - 1}$, but also
for $z \in \{\frac{n}{N}\}_{n= 0, \ldots, N - 1}
\cup \{\frac{n}{N- 1}\}_{n = 0, \ldots, N - 2} \cup \{\frac{n}{N- 2}\}_{n = 0, \ldots, N - 3}
\cup \ldots \{\frac{n}{2}\}_{n = 0, 1}$.
In this case, we can reduce the number of unknowns by one on every
step of the recursion, as we have in the algorithm for calculating
$({\cal F}^D)^{- 1}$.
The complexity of such a modified algorithm is
\[
\sum_{k = 1}^{N} k \log{k} = \log{(H(N)} = \log{ \prod_{k = 1}^N k^k},
\]
where $H$ denotes the {\it hyperfactorial} function. This modification is an improvement.
For $N = 1000$, the difference between the complexities of the modified and unmodified
is roughly $3,7 \times 10^6$.

State of the art numerical methods concerning the nonlinear Fourier transform can be found in
\cite{Osborne} and the references therein.

Our discretizations do not stem from some integrable
discretisation of the integrable systems. However, in our opinion, they shed light on the structure
of nonlinear Fourier analysis. Also, nonintegrable tools are often used in the
study of integrable systems, see \cite{DauxPey},  \cite{LaserRadiation} and many other references.

In a forthcoming paper, we intend to investigate what the results of this paper can tell us about the continuous nonlinear Fourier transformation and its applications in nonlinear dynamics. We also intend
to explore the usefulness of  ${\cal F}^D$ and $({\cal F}^D)^{- 1}$ in the nonlinear Fourier analysis of 
distributions.

\section{Acknowledgement} 
The research for this paper was supported in part by the research project, ARRS: J1-3005
from ARRS, Republic of Slovenia.

\end{document}